\documentclass[graybox]{svmult}

\usepackage{type1cm}        
\usepackage{makeidx}         
\usepackage{graphicx}        
\usepackage{multicol}        
\usepackage[bottom]{footmisc}

\usepackage{newtxtext}       %
\usepackage[varvw]{newtxmath}       
\usepackage{amsmath}
\usepackage{amsthm}

\usepackage{latexsym}
\usepackage{amssymb}
\usepackage{bm}
\usepackage{bbm}
\usepackage{appendix}
\usepackage{graphics,epstopdf}
\usepackage{color}
\usepackage[colorlinks=true,linkcolor=blue,citecolor=red,plainpages=false,pdfpagelabels]{hyperref}
\usepackage[normalem]{ulem}
\usepackage{newlfont}
\usepackage{amsfonts}
\usepackage{amsthm}
\usepackage{graphicx}
\usepackage{epsfig}
\usepackage{amssymb,amsmath,graphicx,amsfonts,color,epsfig,amsxtra,amstext,latexsym,hyperref,setspace,amsthm,array,braket}
\usepackage[thinc]{esdiff}
\usepackage{float}

\usepackage{times}

\makeindex             

\begin{document}

\title*{Black hole thermodynamics in modified gravity }

\author{ Mriganka Dutta\orcidID{0009-0000-3694-9240}\\ Banibrata Mukhopadhyay\orcidID{0000-0002-3020-9513}}

\institute{ Mriganka Dutta \at Department of Physics, Indian Institute of Science, Bangalore 560012, India \email{mrigankad@iisc.ac.in}\and Banibrata Mukhopadhyay \at Department of Physics, Indian Institute of Science, Bangalore 560012, India \email{bm@iisc.ac.in}}

%
%
\maketitle
\vspace{-2cm}
\textit{To be published in Astrophysics and Space Science Proceedings, titled "The Relativistic Universe: From Classical to Quantum, Proceedings of the International Symposium on Recent Developments in Relativistic Astrophysics", Gangtok, December 11-13, 2023: to felicitate Prof. Banibrata Mukhopadhyay on his 50th Birth Anniversary", Editors: S Ghosh \& A R Rao, Springer Nature}\\

\abstract{The theory of general relativity is often considered under the framework of modified Einstein gravity to explain different phenomena under strong curvature. The strong curvature effect plays a main role near black holes, where the gravitational field is strongest. The idea of black hole thermodynamics is to describe the strong field curvature properties of a black hole in the effective thermodynamical framework, e.g. entropy, temperature, heat capacity etc. In this paper, our aim is to explore how the effect of modified gravity changes the thermodynamic properties of black hole. We show that even a small modification to Einstein gravity affects the thermodynamical properties of a black hole.}

\section{Introduction}
\label{Sec.1}
The theory of general relativity is one of the finest and most successful theories that we have, for over 100 years, to describe the spacetime structure. The theory explains how mass affects the spacetime curvature and in turn how spacetime curvature affects the path of mass transportation. Einstein's general relativity has been verified by many phenomena such as gravitational lensing \cite{Wambsganss:1998gg}, perihelion precession of Mercury's orbit \cite{Kraniotis:2003ig}, gravitational wave generation \cite{LIGOScientific:2016aoc}, etc. However, most of these are done based on the weak field approximation of general relativity. It is still uncertain whether the theory holds its validity in the presence of strong curvature in spacetime.  

Einstein's gravity theory also predicts that if we have concentrated mass at a very small volume, it can form a one-way membrane in spacetime. This membrane is called the event horizon surface. If classically, whatever crosses the event horizon getting trapped inside it, we call that structure black hole. Karl Schwarzschild gave the spacetime metric structure for non-rotating and later Roy Kerr deduced the spacetime metric structure for rotating black holes.
Spacetime near black holes is such a strong field regime that the validity of general relativity remains uncertain. Hence one may need to consider the modification of gravity theory.

In 1971, Roger Penrose showed that theoretically that it should be possible to extract energy from a rotating black hole. Consequently, it was understood that the area of a black hole (i.e. the area of the event horizon surface) never decreases. Based on these ideas, Jacob Bekenstein further formulated a connection between the black hole area and its entropy \cite{Bekenstein:1973ur}. He also proposed the idea of the generalized second law of thermodynamics (GSL). Later in 1975, Steven Hawking showed that due to quantum mechanical effects, black holes can radiate particles and have some finite temperature \cite{Hawking:1975vcx}. All of these works are done assuming Einstein's gravity theory holds well in strong spacetime curvature. 

One of the frameworks of modified gravity theory is that
asymptotically, hence in a weaker spacetime curvature, a modified gravity leads to the results of Einstein's gravity theory. However, the modified gravity has significant variations in thermodynamic properties near black holes. In this work, we shall try to explain what happens to the thermodynamic properties of black holes in the presence of modified gravity.

\section{Metric of rotating black hole in modified gravity }
\label{Sec.2}
The modification to general relativity can be done in various ways. Out of various models of modified gravity, we consider the $f(R)$-gravity. The idea is to have asymptotically flat solutions such that on a large scale, the modified gravity behaves as regular gravity theory. The Einstein-Hilbert action for a $f(R)$-gravity (with metric signature [+,-,-,-] ) is
\begin{equation}
    S=\int \left(\frac{c^4}{16\pi G} f(R) +\mathcal{L}_{M} \right)\sqrt{-g}~ d^4x.
    \label{1}
\end{equation}
Here `$-g$' is the determinant of the metric tensor $g_{\mu\nu}$, $\mathcal{L}_M $ is the Lagrangian density of matter, $f(R)$ is a function of Ricci scalar $R$. The expression for $f(R)$ has to be such that $df(R)/dR$ behaves as $1+B/r$ so that at large $r$ the gravity theory acts as the Einstein's gravity theory. 

 Following a previous work \cite{Das:2022xih},
one can consider the metric to describe the rotating black hole spacetime in a modified gravity.
We take the metric tensor, in ($t,r,\theta,\phi$) coordinates, as
\begin{equation}
    g_{\mu\nu}=
    \begin{pmatrix}
        1-\frac{2Mr}{\Sigma}-\frac{\beta}{\Sigma} & 0 & 0 &\frac{Ma\sin^2{\theta}}{\Sigma}(2r+\beta)  \\
        0 & -\frac{\Sigma}{\Delta} &0 &0\\
        0 & 0 & -\Sigma & 0\\
        \frac{Ma\sin^2{\theta}}{\Sigma}(2r+\beta) & 0 &0 &- \sin^2\theta\left(r^2+ a^2 +\frac{a^2 \sin^2{\theta}(2Mr+\beta)}{\Sigma}\right)
    \end{pmatrix},
    \label{2}
\end{equation}
  where $\Sigma=r^2 + a^2\cos^2{\theta}$, $a$ is the specific angular momentum of black hole, $\beta$ is $B(B-6)/2$ and $\Delta=\Sigma ~ g_{tt}/X(r,\theta)+a^2\sin^2{\theta}$. The expression for $X(r,\theta)$ is given as $(r^2 + a^2\cos^2{\theta})^2/[(r+B/2)^2 + a^2\cos^2{\theta}]^2$. It needs to be stated that the value of $B$ has to be negative and we restrict the modification such that $|B|<1$. For $B>0$ we have unphysical properties of gravity. Considering the fact that $|B|$ is small; we can take $X(r,\theta)\approx 1$.
\section{Thermodynamic properties}
\label{Sec.3}
\subsection{Temperature and entropy of black hole in modified gravity}
\label{Sec.31}
The temperature of a black hole can be calculated from the surface gravity, once the line element or the metric tensor of spacetime is determined. For the metric, mentioned in Eq. (\ref{2}), we have two Killing vectors. One for the time-translation in-variance and another for rotational in-variance along $\phi$ direction.
Thus for a rotating black hole the Killing vector can be taken as 
\begin{equation}
    K^\mu=[1,0,0,\Omega_H], 
\label{3}
\end{equation}
where $\Omega_H=-g_{t\phi}/g_{\phi\phi}$ at the horizon. To determine the horizon we make $g_{rr}\rightarrow\infty$, i.e. $\Delta$ to be 0. Considering $|B|<1$, $\Delta$ can be simplified to $r^2-2Mr+a^2-\beta$. This gives the radii of the outer and inner horizons as 
\begin{equation}
r_{\pm} = M\pm\sqrt{M^2-a^2+\beta},
\label{4}
\end{equation}
where $r_+$ and $r_-$ denote the outer and inner horizons respectively. The area corresponding to $r_+$ turns out to be 
\begin{equation}
    A=4\pi(r_{+}^2+a^2).
    \label{5}
\end{equation}
\begin{figure}
    \centering
    \includegraphics[scale=0.7]{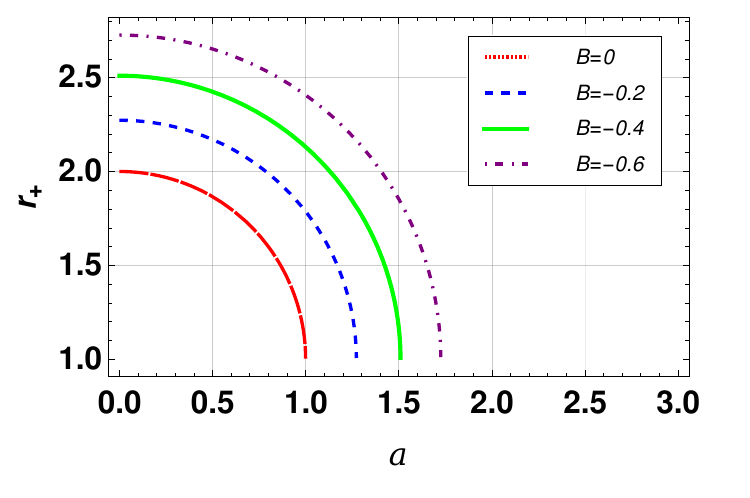}
    \caption{Radius of the outer horizon as a function of spin parameter, for different values of $B$ parameter considering $M=1$.}
    \label{Fig.1}
\end{figure}
The surface gravity of a black hole ($\kappa$) can be calculated from the Killing vector by 
\begin{equation}
    \kappa^2=\lim_{r\rightarrow r_+} \frac{(\nabla_{\nu}(K^{\mu}K_{\mu})\nabla^{\nu}(K^{\mu}K_{\mu}))}{4(K^{\mu}K_{\mu})}.
\label{6}
\end{equation}
From the surface gravity one can define temperature of black hole as $T_{BH}^{'}=\kappa/(2\pi)$. It turns out that, one can also define the temperature of the black hole just by differentiating the area $A$ \cite{Bekenstein:1973ur}. After differentiation we obtain
\begin{equation}
    dM=\frac{r_+-r_-}{16\pi(r_{+}^2+a^2)} dA+ \frac{a}{(r_{+}^2+a^2)}dL,
    \label{7}
\end{equation}
where black hole mass $M$, area $A$ and total angular momentum $L$ (=$aM$) are taken to be independent variables. The Eq. ({\ref{7}}) gives the essence of the second law of thermodynamics for a black hole. Following the earlier works of Bekenstein  \cite{Bekenstein:1973ur} and Hawking  \cite{Hawking:1975vcx}, the relation between the area and entropy of a black hole is $S=A/4$. 
\begin{figure}
    \centering
    \includegraphics[scale=0.7]{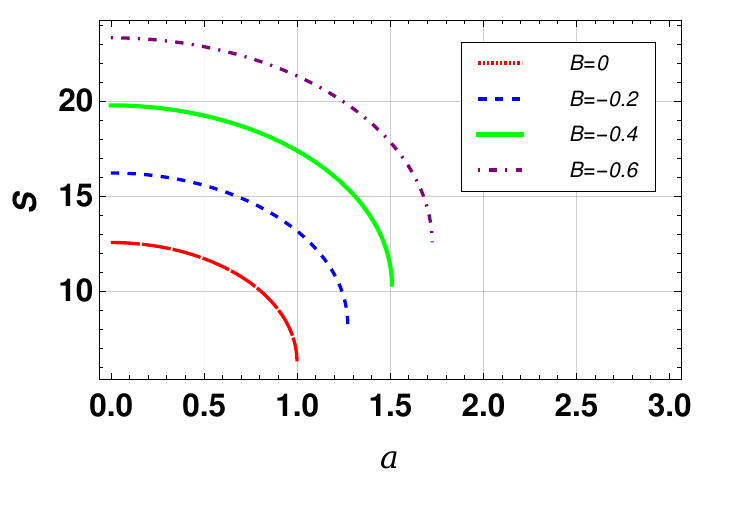}
    \caption{Entropy of the black hole as a function of spin parameter, for different values of $B$ parameter considering $M=1$.}
    \label{Fig.2}
\end{figure}
Hence one obtains the temperature from Eq. ({\ref{7}}) as 
\begin{equation}
    T_{BH}=\frac{r_+-r_-}{4\pi(r_{+}^2+a^2)}.
    \label{8}
\end{equation}
It can be verified that the temperature expressions we obtain from Eq. (\ref{6}) and Eq. (\ref{8}) are same, i.e. $T_{BH}^{'} =T_{BH}$.
\begin{figure}[t]
    \centering
    \includegraphics[scale=0.7]{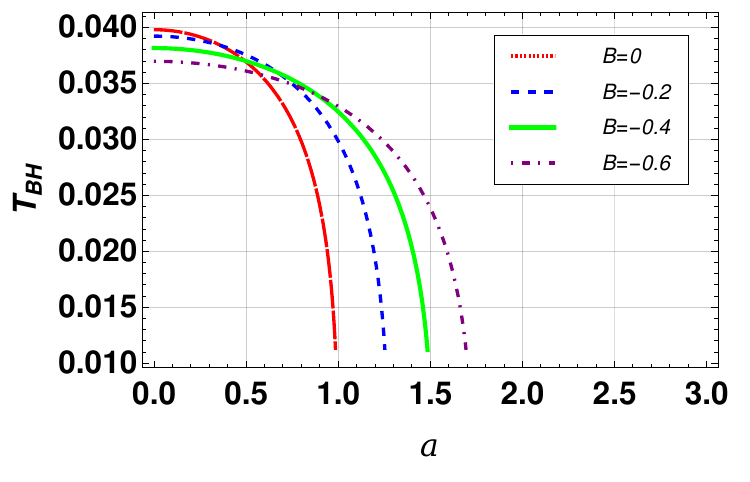}
    \caption{Temperature of the black hole as a function of spin parameter($a$) for different values of $B$ parameter considering $M=1$.}
    \label{Fig.3}
\end{figure}
 
\subsection{Heat capacity of black hole in modified gravity}
\label{Sec.32}
Following the thermodynamic relations we can say that the `heat capacity' \cite{Davies:1978zz} of black hole is defined as 
\begin{equation}
    C_L=T\left(\frac{\partial S}{\partial T}\right)_L.
    \label{9}
\end{equation}
To obtain an analytical form of heat capacity, one can define $M$ as a function of $S,T_{BH}$ and $L$. From the Eq. (\ref{5}), the mass of black hole can be written as
\begin{equation}
    M=\left(\frac{S}{4\pi}+\frac{\pi\beta^2}{4S}+\frac{4\pi L^2}{S}-\frac{\beta}{2}\right)^{1/2}.
    \label{10}
\end{equation}
Thus temperature in terms of entropy, mass and angular momentum  becomes
\begin{equation}
    T_{BH}=\frac{1}{2M}\left(\frac{1}{4\pi}-\frac{\frac{\beta^2}{4}+L^2}{S^2}\right).
    \label{11}
\end{equation}
Now using Eq. (\ref{11}) and Eq. (\ref{10}) in Eq. (\ref{9}) we obtain `heat capacity'
\begin{equation}
    C_L=\frac{MT_{BH}S}{\frac{1}{4\pi}-2MT_{BH}-T_{BH}^2S}.
    \label{12}
\end{equation}
\begin{figure}
    \centering
    \includegraphics[scale=0.7]{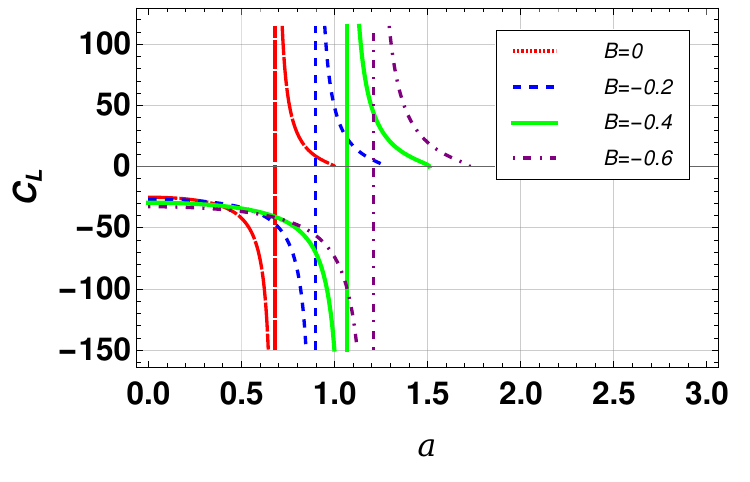}
    \caption{Heat capacity at constant angular momentum of the black hole as a function of the spin parameter, for different values of $B$ parameter considering $M=1$.}
    \label{Fig.4}
\end{figure}
\begin{figure}[H]
    \centering
    \includegraphics[scale=0.7]{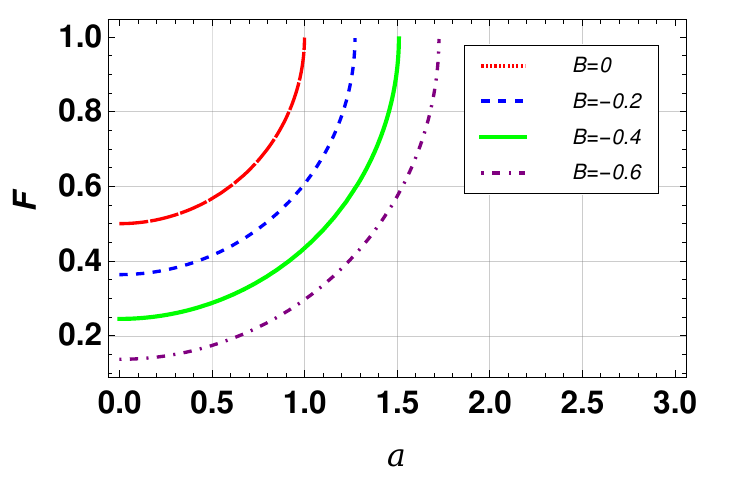}
    \caption{Free energy of the black hole as a function of the spin parameter, for different values of $B$ parameter considering $M=1$.}
    \label{Fig.5}
\end{figure}

\subsection{Free energy of black hole in modified gravity}
\label{33}
The free energy of a thermodynamic system is described as the energy available for work done. Similarly considering the thermodynamic properties of black holes, one can define the free energy. Free energy of a black hole determines how strong the spacetime curvature effect is. We define the free energy of black holes as
\begin{equation}
    F=M-T_{BH}S.
    \label{13}
\end{equation}
To obtain the entropy of a black hole, as a function of $M$, $L$ and $\beta~(=B(B-6)/2$), we write Eq. (\ref{5}) as
\begin{equation}
    S=2\pi(M^2+\beta/2)+2\pi\sqrt{M^4+M^2\beta-L^2}.
    \label{14}
\end{equation}

\section{Implications}
\label{4}

We have considered the metric for rotating black holes in a modified gravity. From Fig. {\ref{Fig.1}} we see that with decreasing $B$ there is an increase of $\beta$ and hence the increase of outer horizon radius, at a fixed specific angular momentum $a$. The entropy, as in Fig. {\ref{Fig.2}}, of the black hole also increases as the outer horizon radius increases. The behavior of heat capacity (Fig. {\ref{Fig.4}}) shows a sudden change from a negative value to a positive value, at some constant specific angular momentum, i.e. a phase transition sets in. The value of specific angular momentum, where this phase transition takes place, shifts towards higher values of specific angular momentum with decreasing $B$. This effect is seen because the effect of modified gravity opposes the effect of angular momentum, hence supporting a higher value of $a$. 
The free energy of a black hole is also a thermodynamic property that is drastically affected by modified gravity. From Fig. {\ref{Fig.5}} we see that with decreasing $B$ or increasing $\beta$, the free energy decreases. The lower value of the free energy makes the curvature of spacetime less affected. The lesser curvature effect causes the outer horizon radius to increase.  

\section{Conclusions}
\label{5}
We have explored a $f(R)$ gravity in a strong gravitational field regime. Considering the modified metric for a rotating black hole, we obtain analytical expressions for temperature, entropy, heat capacity, and free energy. This work shows significant variation in the thermodynamic properties compared to those in Einstein's gravity. Thus one can not ignore the possibility of a modification of gravity theory, particularly to probe properties of spacetime close to the black hole. As a future work, we plan to look for the modification of black hole evaporation considering modified gravity theory, see \cite{PB} in this volume, for a preliminary result.

\begin{acknowledgement} MD thanks the organizers of the International Symposium on Recent Developments in Relativistic Astrophysics (ISRA) for giving the opportunity to write this paper in this special volume.

\end{acknowledgement}

\end{document}